\documentclass[pra,twocolumn,showpacs,floatfix]{revtex4}

\usepackage{graphicx}
\usepackage{amsmath}
\usepackage{amssymb}
\usepackage{epsf,latexsym}
\usepackage{times}
\usepackage{bm}

\usepackage{float}

\newcommand{\ket}[1]{\left| #1 \right\rangle}
\newcommand{\bra}[1]{\left\langle #1 \right|}

\newcommand{\be}{\begin{equation}}
\newcommand{\ee}{\end{equation}}
\newcommand{\ba}{\begin{eqnarray}}
\newcommand{\ea}{\end{eqnarray}}

\newcommand{\ua}{\uparrow}
\newcommand{\da}{\downarrow}

\begin{document}
\title{The Hubbard model as an approximation to the entanglement in nanostructures}
\author{J. P. Coe$^{1}$
}
\email{jpc503@york.ac.uk}
\author{V. V. Fran\c{c}a$^{2}$
}
\email{vivian.franca@physik.uni-freiburg.de}
\author{I. D'Amico$^{1}$
}
\email{ida500@york.ac.uk}
\affiliation{
$^1$ Department of Physics, University of York, York YO10 5DD, United Kingdom.\\
$^2$ Physikalisches Institut, Albert-Ludwigs-Universit\"{a}t, Hermann-Herder-Stra\ss e 3, D-79104 Freiburg, Germany.}
\date{\today}

\begin{abstract}
We investigate how well the one-dimensional Hubbard model describes the entanglement of particles trapped in a string of quantum wells.   We calculate the  average single-site entanglement for two particles interacting via a contact interaction and consider the effect of varying the interaction strength and the interwell distance.
We compare the results with the ones obtained within the one-dimensional Hubbard model with on-site interaction.
We suggest an upper bound for the average single-site entanglement for two electrons in $M$ wells and discuss analytical limits for very large repulsive and attractive interactions. We investigate how the interplay between interaction and potential shape in the quantum well system dictates the position and size of the entanglement maxima and the agreement with the theoretical limits. 
Finally we calculate the spatial entanglement for the quantum well system and compare it to its average  single-site entanglement.
\end{abstract}

\maketitle

\section{Introduction}
Entanglement is considered one of the main resources in quantum information and a reason why quantum computers may be used for computing feats that could not be achieved with traditional processors \cite{NIELSEN}. In this context  quantum dots are thought of as a viable possibility in the quest to construct scalable quantum processors \cite{CHEN00,hodgson:114319,FENG04,SPILLER07,PhysRevLett.83.4204,XIAOQIN03,PhysRevA.57.120,IDA3}. With this in mind, finding accurate ways to calculate the entanglement between electrons in quantum dots becomes important for quantum information processing.  However modeling these many-body quantum systems often necessitates the employment of approximations.  For example, one-dimensional wells may be used in the study of spherically symmetric quantum dots and aid understanding of trends in more general quantum dot systems \cite{Abdullah}. 
 
The  Hubbard model \cite{OrigHubbard} allows interacting many-body systems to be simulated by mapping them onto a lattice model with (usually) on-site interactions only. Despite its relative simplicity it still captures a significant amount of physics: for example, in strongly correlated fermionic systems it has been used to model particles trapped in an optical lattice \cite{Xianlong06}, high $T_{c}$ superconductivity \cite{Machida20069}, and the metal-insulator transition \cite{PhysRevB.58.9607}. The one-dimensional homogeneous Hubbard model (HM) also benefits from the existence of an exact solution in the thermodynamic limit \cite{LiebWu}.  Recently the use of the Hubbard model as an approximation to the exchange coupling in quantum-dot nanostructures has been investigated \cite{PEDERSEN07}.  The entanglement of the one-dimensional Hubbard model has been investigated in Refs.~\cite{PhysRevLett.93.086402,VIVIAN,Larsson05}.  A local-density approximation (LDA) to the entanglement has been proposed in Ref.~\cite{VIVIAN1} and applied to inhomogeneous systems. 

In this paper we compare the Hubbard model predictions to results from a system of two interacting fermions trapped within a chain of square well potentials. Here each well corresponds to one of the Hubbard model sites. We consider both repulsive (electron-electron) and attractive interactions, and calculate the corresponding entanglement when the strength of the interaction, the chain length, and the interwell distance are changed. We compare these results with the average single-site entanglement calculated from the Hubbard model.

In doing this we infer information on the accuracy of using results from the Hubbard model to approximate the average single-site entanglement of the quantum well system.  If the entanglement of electrons in quantum wells can be described using the Hubbard model, then,  by using the powerful LDA formalism developed in \cite{VIVIAN1}, we could in principle calculate the entanglement present in quantum well systems with a large number of interacting electrons. This would be a non-trivial result as a direct calculation of the entanglement for a system with a large number of interacting electrons becomes computationally prohibitive as the number of particles increases. 

In Sec.~\ref{sec:Model}  we introduce the one-dimensional quantum well system and discuss how we numerically calculate the single-site entanglement.  In Sec.~\ref{sec:deltaHubbard} we compare the results from the Hubbard model  with the ones from the electron system for different interwell distances, chain length, and  Coulomb interaction strength.  We also investigate how the interaction strength affects the electronic density, and its effect on the matching between the Hubbard and quantum well system. In Sec.\ref{sec:chap6maxS} we propose an upper bound for the average single-site entanglement of the quantum well system  and discuss the large Coulomb interaction limit. We investigate how close our numerical results come to  these analytical expressions. In Sec.~\ref{attr-vs-rep} we compare attractive and repulsive particle-particle interaction, as well as discussing the large interparticle attraction limit. 
Sec.~\ref{sec:spatial} is devoted to the comparison between the average single-site entanglement with the spatial entanglement for the quantum well system and finally    Sec.~\ref{sec:Conclusions} contains our conclusions.

\section{The quantum well two-electron system}
\label{sec:Model}
The Hamiltonian, in atomic units, for the one-dimensional system of two electrons confined within an array of quantum wells (QWs) is given by
\begin{equation}
H=\sum_{i=1,2}\left(-\frac{1}{2}\frac{d^{2}}{dx_{i}^{2}}+v(x_{i}) \right )+C_{U}f(\left|x_{1}-x_{2}\right|).
\label{H_wells}
\end{equation}
 The potential $v(x_{i})$ models a string of regularly spaced, identical square wells, symmetric around the origin, and defined by the quantities:  $M$ the number of wells,  $w$ the width of each well, $d$ the barrier width between two consecutive wells, and $v_{0}$ the depth of each well.  We set $f(x)=\delta(x)$ to model a contact Coulomb repulsion and define $C_{U}$ as the interaction strength.  $C_{U}$ will allow us to compare the system with the Hubbard model.

We solve the time independent Schr\"{o}dinger equation corresponding to Eq.~(\ref{H_wells}) by using `exact' diagonalization; the electronic ground state is a singlet thereby satisfying the stipulation of zero magnetization. We calculate the spatial part of the many-body wavefunction using the first $N$ eigenfunctions of the potential $v(x)$ as single-particle basis functions.  We employ these to produce a basis of  symmetric two-particle wave-functions which means we only need to consider $N(N+1)/2$ functions.  The form of the Hamiltonian is conducive to this method as by varying the interaction strength independently we do not need to recalculate the basis, or any integral involved in the diagonalization, as could be the case if we varied the well geometry directly. In this respect we note that  a system with $C_{U}=K$, depth $v_{0}$, well width $w$ and barrier width $d$ is equivalent to a system  with $C_{U}=1$, depth $v_{0}/K^{2}$, well width $Kw$, and barrier width $Kd$.

\subsection{Average single-site entanglement}
We wish to calculate the average single-site (or local) entanglement of the system ground state. This type of entanglement is relevant for systems of indistinguishable fermions \cite{ZANARDI}. To this aim we divide our QW system in contiguous `sites', each site centered around a single well.  

The entanglement entropy $S$ of the system is given by 
\begin{equation}
S=\frac{1}{M}\sum_{i}^{M}S_{i}, 
\label{eq:avsinglesite}
\end{equation}
with 
\begin{equation}
S_{i}=-Tr \rho_{red,i} \log_{2} \rho_{red,i}
\label{eq:singlesite}
\end{equation}
the {\it i}-site von Neumann entropy of the reduced density matrix $\rho_{red,i}$.
The von Neumann entropy is considered as one of the definitive measures of entanglement for a pure bipartite system.  By dividing the system into sites and moving to a site-occupation basis the reduced density matrix $\rho_{red,i}$  becomes a $4\times 4$ diagonal matrix \cite{ZANARDI,Larsson06},
\begin{equation}
\rho_{red,i}=\text{diag}\left[P_i(\uparrow\downarrow),P_i(\uparrow),P_i(\downarrow),P_i(0)\right]
\end{equation}
with $P_i(\gamma)$ the probability of double ($\gamma=\uparrow\downarrow$), single ($\gamma=\uparrow$ or $\downarrow$), or zero ($\gamma=0$) electronic occupation at site $i$ \cite{Larsson05}.   

We calculate the ground-state wave-function, for an even number $M$ of wells, and from that obtain the occupation probabilities.  We calculate the probability that both electrons are in the same left-most ($M$th) well 
as \begin{equation}
P_{M}(\uparrow\downarrow)=\int_{-r_{c}}^{b}\int_{-r_{c}}^{b} |\psi(x_{1},x_{2})|^{2} dx_{1}dx_{2},
\end{equation} where $r_{c}$ is the (numerical) integration cut-off point and $b=-(M/2-1)(w+d)$  the mid point between the left-most well and the next well.

The probability that only one spin up (or spin down by symmetry) electron is in this well is
\begin{equation}
P_{M}(\uparrow)=P_{M}(\downarrow)=\int_{-r_{c}}^{b}\int_{b}^{r_{c}} |\psi(x_{1},x_{2})|^{2} dx_{1}dx_{2}.
\end{equation}
$P_M(0)$ may then be deduced, as the probabilities sum to $1$.

For the other wells the occupation probabilities are
\begin{equation}
P_{j+M/2}(\uparrow\downarrow)=\int_{a}^{b}\int_{a}^{b} |\psi(x_{1},x_{2})|^{2} dx_{1}dx_{2},
\end{equation}
and
\begin{eqnarray}
P_{j+M/2}(\uparrow)=\int_{a}^{b}\int_{-r_{c}}^{a} |\psi(x_{1},x_{2})|^{2} dx_{1}dx_{2}\\
+\int_{a}^{b}\int_{b}^{r_{c}} |\psi(x_{1},x_{2})|^{2} dx_{1}dx_{2},
\end{eqnarray} with
$a=-j(w+d)$ and $b= -j(w+d)+d+w$, $1 \leq j \leq (M/2-1)$. 
As we only consider an even number of wells distributed symmetrically about the origin, the probability values for wells $M/2$ to $1$ are known by symmetry.

\section{Comparison with the Hubbard model}
\label{sec:deltaHubbard}
The Hubbard model \cite{OrigHubbard} is described by the Hamiltonian 
\begin{eqnarray}
\nonumber H_{\text{HM}}=-t\sum_{i,\sigma}\left (c_{i,\sigma}^{\dagger}c_{i+1,\sigma}+c_{i+1,\sigma}^{\dagger}c_{i,\sigma} \right)\\
+\tilde{U}\sum_{i}\hat{n}_{i,\uparrow}\hat{n}_{i,\downarrow},
\label{eqn:HubbardHamiltonian}
\end{eqnarray}
where $i$ runs over the $M$ sites and $\sigma=\uparrow,\downarrow$.  Here $t$ is the hopping parameter and $\tilde{U}$ is the interaction strength.  $c_{i,\sigma}^{\dagger}$ ($c_{i,\sigma}$)  creates (destroys) a particle of spin $\sigma$ at site $i$ while $\hat{n}_{i,\sigma}=c_{i,\sigma}^{\dagger}c_{i,\sigma}$ is the particle number operator. We solve Eq.~(\ref{eqn:HubbardHamiltonian}) by exact diagonalization in the single-site occupation basis $\{\ket{\uparrow \downarrow},\ket{\uparrow},\ket{\downarrow}, \ket{0}\}$. We apply open boundary conditions and consider an average particle density of $n=n_{\downarrow}+n_{\uparrow}=2/M$, with $n_{\sigma}$ the average density of the $\sigma$-spin component. 
Again we calculate the average single-site entanglement Eq.~(\ref{eq:avsinglesite}) \cite{PhysRevLett.93.086402,VIVIAN,Larsson05}.  
 Usually the hopping parameter $t$ is used to rescale $\tilde{U}$, giving the dimensionless interaction strength 
$U=\tilde{U}/t$.

To compare results, we need to calculate the equivalent of $t$ and $\tilde{U}$ for the QW system discussed in Sec.~\ref{sec:Model}.  In the Hubbard model  the hopping parameter is the expectation value of the single-particle operators in the Hamiltonian with respect to the single-particle wave functions localized at adjacent sites.  When the hopping parameter $t$ is independent of the sites  it may be written as
\begin{equation}
t=\bra{\phi_{i}(\bm{r})} \left(-\frac{1}{2}\nabla^{2}+V(\bm{r}) \right) \ket{\phi_{i+1}(\bm{r})},
\label{eq:hopping}
\end{equation}
where  $\phi_{i}$ is the wavefunction at any site $i$ and  $V(\bm{r})$ is the single-particle confining potential. 

Following this definition,  we estimate the hopping parameter for our quantum well model as 
\begin{equation}
t_{w}=\bra{\phi_{L}(x)} \left(-\frac{1}{2}\frac{d^{2}}{dx^{2}}+v(x) \right) \ket{\phi_{R}(x)}, \label{t_w}
\end{equation}
where $\phi_{L(R)}$ has the shape of  the single-particle ground state of the finite single square well potential $\phi_{w}$, but centered in the left ($\phi_{L}$) or right ($\phi_{R}$) well.
Here $v(x)$ is the potential defined in Eq.~(\ref{H_wells}), with the zero of energy chosen such that  $v(x)$ has zero as its lowest value thereby ensuring that the potential contribution is always positive and an increase in the depth of the well causes the hopping parameter to decrease.
The phase of $\phi_{L}$ and $\phi_{R}$ is chosen to make $t_{w}$ real and positive.

The on-site interaction in the Hubbard model \cite{OrigHubbard} is defined by
\begin{equation}
\tilde{U}=\frac{1}{2}\bra{\phi_{i}(\bm{r}_{2})}\bra{\phi_{i}(\bm{r}_{1})} \frac{1}{|\bm{r}_{1}-\bm{r}_{2}|} \ket{\phi_{i}(\bm{r}_{1})}\ket{\phi_{i}(\bm{r}_{2})}.
\end{equation}
The corresponding parameter in our $1$D model with a delta function interaction is then
\begin{equation}
\tilde{U}_{w}=\frac{C_{U}}{2}\int \phi_{w}^{4}(x) dx\label{U_w}.
\end{equation}

We may now estimate $U$ for our model as
\begin{equation}
U\approx\frac{\tilde{U}_{w}}{t_{w}}.
\label{eq:Udefn}
\end{equation}
For systems of wells characterized by the parameters $w=d=2$ $a_{0}$, where $a_{0}$ is the (effective) Bohr radius, and $v_{0}=10$ (effective) Hartree, we find $U=12344C_{U}$; when the interwell distance is reduced to $d=0.2$ $a_{0}$ we obtain  $U=3.14C_{U}$, and $U=1.27C_{U}$ for the limiting case $d=0$.

\subsection{Effect of many-body interactions on the electronic density}
\label{sec:chap6varywells}
\begin{figure}[ht]\centering
  \includegraphics[width=.4\textwidth]{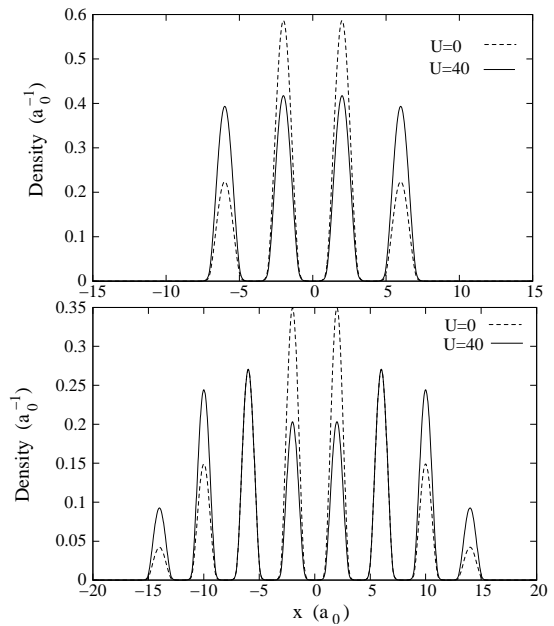}
  \caption{Upper Panel: electron density for $U=0$ ($C_{U}=0$) (dashed line) and  $U=40$ ($C_{U}=0.00324$) (solid line) for a  $4$ well potential and $w=2$ $a_{0}$, $d=2$ $a_{0}$ and $v_{0}=10$ Hartree. Lower Panel: as for the upper panel but for a system of $8$ wells. }\label{fig:Deltawelldensityd=2}
\end{figure}

\begin{figure}[ht]\centering
  \includegraphics[width=.4\textwidth]{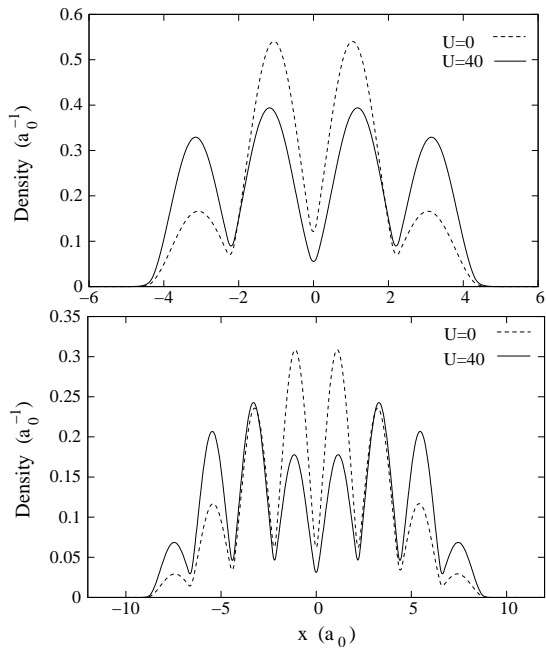}
  \caption{Upper panel: electron density for $U=0$ ($C_{U}=0$) (dashed line) and  $U=40$ ($C_{U}=12.75$) (solid line) for a $4$ well potential with $w=2$ $a_{0}$, $d=0.2$ $a_{0}$ and $v_{0}=10$ Hartree.  Lower Panel: as for the upper panel but for a system of $8$ wells.}\label{fig:Deltawelldensityd=0.2}
\end{figure}

Next we explore how the electron density alters to maximize its exposure to the attractive confining potential whilst attempting to minimize the interaction between the electrons. For two wells the density profile is clearly symmetric so we do not discuss it further.  A non-interacting four well system (upper panel of Fig.~\ref{fig:Deltawelldensityd=2}, dashed line) displays a density clearly higher in the inner wells. However for  $U=40$, due to the electron repulsion, the difference between the electron density in the inner and outer wells becomes much smaller  (solid line).
For eight wells and $U=0$ the inner wells are again preferentially occupied while there is very little density in the outermost wells (Fig.~\ref{fig:Deltawelldensityd=2}, lower panel, dashed line).  When the interaction is `switched on' to $U=40$ the two central wells have a lower peak density than the nearby wells to compensate for the Coulomb repulsion,  but the outermost wells still display, by comparison, a much lower density.   Similar behaviors of the interacting and non-interacting densities are found when the distance between wells is reduced by an order of magnitude ($d=0.2$ $a_{0}$). However in this case the density is considerably different from zero in the barrier region (Fig.~\ref{fig:Deltawelldensityd=0.2}).

These results seemingly show that, apart from accidental compensation, the electron density in the different wells can not be made equal by applying an unmodulated Coulomb interaction.

\subsection{Comparison of  entanglement results}

\begin{figure}[ht]\centering
  \includegraphics[width=.4\textwidth]{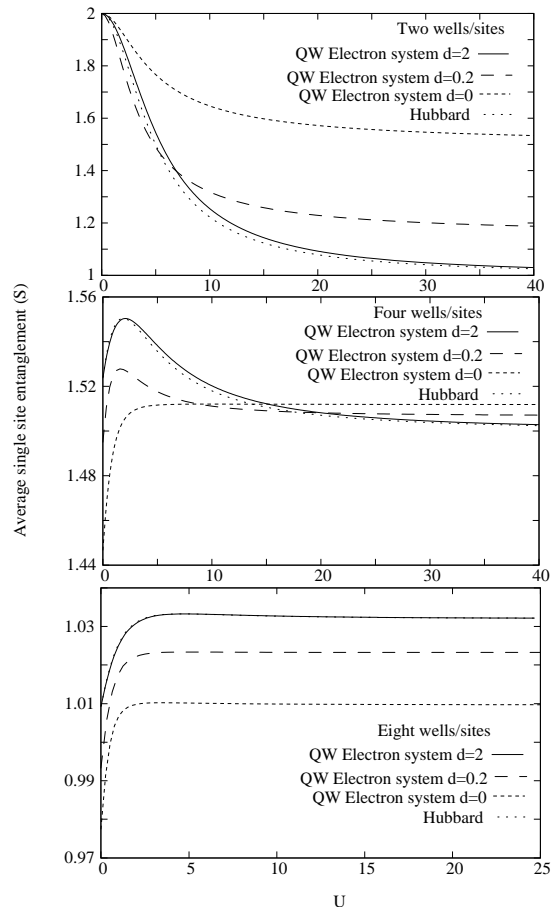}
  \caption{ Average single-site entanglement for the Hubbard model and the QW electron system with $U=\tilde{U}_{w}/t_{w}$, $w=2$ $a_{0}$, $d=2$ $a_{0}$, $d=0.2$ $a_{0}$ or $d=0$, and $v_{0}=10$ Hartree.  Upper panel: $2$ sites with $n=1$ $a_{0}^{-1}$ and $2$ wells. Center panel: $4$ sites with $n=0.5$ $a_{0}^{-1}$ and $4$ wells.  Lower panel: $8$ sites with $n=0.25$ $a_{0}^{-1}$ and $8$ wells.}\label{fig:ContactGraphsApproxt}
\end{figure}

 In Fig.~\ref{fig:ContactGraphsApproxt} we compare the average single-site entanglement for the Hubbard model with that of the QW electron systems characterized by  $d=2$, $d=0.2$, and the limiting case $d=0$. The latter corresponds to the arbitrary partition of a single well of width $Mw$ into $M$ equal regions.

For $d=2$ $a_{0}$ and two wells (upper panel) we see that the entanglement decreases similarly to the Hubbard model as $U$ increases but the entanglement in the QW system is slightly higher.  When we consider four wells (middle panel), in both cases the entanglement increases up to a maximum and then decreases, the maximum occurring at slightly different values of $U$.  For eight wells (lower panel) the entanglement in the two systems is almost indistinguishable.

For  $d=0.2$ $a_{0}$, two wells, and  $U\lesssim 8$  the Hubbard model is fairly accurate  in reproducing the average single-site entanglement, while,
for stronger interactions, results from the Hubbard model reproduce only the qualitative trend (Fig.\ref{fig:ContactGraphsApproxt}, upper panel). For $U\gtrsim 8$ the entanglement values are intermediate between the Hubbard model and the limiting case $d=0$: in this respect we note that for $d=0.2$ $a_{0}$ even when the repulsive interaction is as high as $U=40$ ($C_{U}=12.75$) the electron density in the QW system does not become localized within the wells  (see Fig.~\ref{fig:Deltawelldensityd=0.2}).

When four wells are considered, the Hubbard model reproduces the entanglement trend qualitatively and is less accurate when the interaction is low (Fig.~\ref{fig:ContactGraphsApproxt}, middle panel). The maximum entanglement is lower for $d=0.2$ $a_{0}$ and in general the entanglement trend is intermediate between the Hubbard model and the limiting case $d=0$.  
For four wells the difference between the maximum values of the entanglement can not be removed by rescaling $U$ (see discussion in next section). 

For eight wells the Hubbard model reproduces the qualitative behavior but the entanglement is lower at all values of $U$ and intermediate with respect to the results for $d=0$.  

It should be noted however that, even for $d=0.2$ $a_{0}$, the percentage error for the entanglement as estimated using the Hubbard model   will be relatively small for the eight and four well systems ($\sim 1\%$), while more substantial for the two-well case and $U \gtrsim 8$ ($\sim 20\%$).

Our results show that the average single-site entanglement of the Hubbard model is a very good match for the entanglement of a QW electron system when wells are far enough apart to prevent significant electron density in the interwell barrier region (see Fig.~\ref{fig:Deltawelldensityd=2}); it is less good, although it gives the general trend, when the wells become closer, as the density profile displays less well-defined `sites' (see Fig.~\ref{fig:Deltawelldensityd=0.2}) .
Surprisingly though, when considering a large number of wells, the Hubbard model reproduces the entanglement within few percent at all interaction strengths, even when compared to the limiting case scenario $d=0$  (Fig.\ref{fig:ContactGraphsApproxt}, lower panel). This results suggests that here the Hubbard model sites could be interpreted as a fine enough mesh discretization of the continuous spatial variable.

\subsection{Rescaling $\tilde{U}_{w}/t_{w}$}
  We now investigate whether, for $d=2$ $a_{0}$, the small discrepancy between the Hubbard model and the QW system results for the entanglement may be removed by choosing an `ad hoc' value of $\tilde{U}_{w}/(t_{w}C_{U})$.

\begin{figure}[ht]\centering
  \includegraphics[width=.4\textwidth]{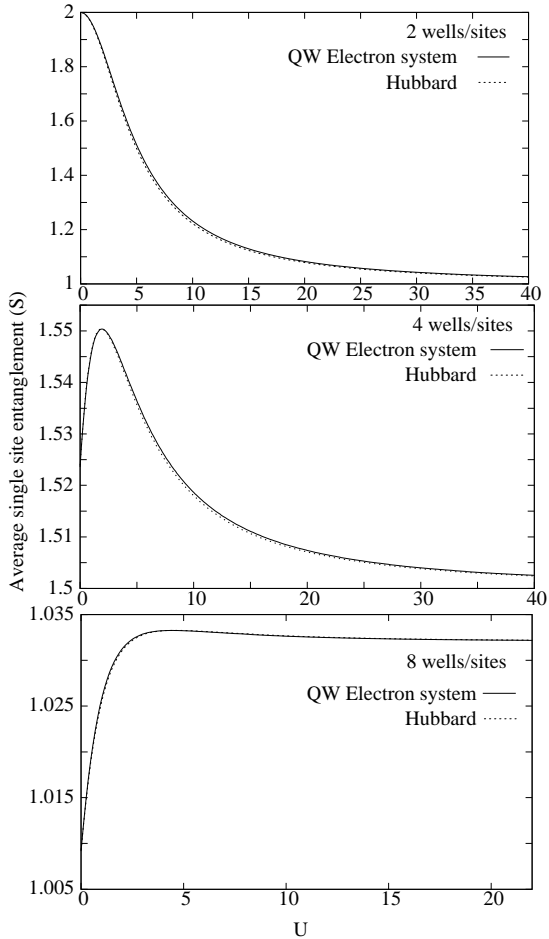}
  \caption{ Average single-site entanglement $S$ for the Hubbard model and the QW electron system with $U=11500C_{U}$, $w=2$ $a_{0}$, $d=2$ $a_{0}$, and $v_{0}=10$ Hartree.  Upper Panel: $2$ sites with $n=1$ $a_{0}^{-1}$ and $2$ wells.  Center Panel: $4$ sites with $n=0.5$ $a_{0}^{-1}$ and $4$ wells.  Lower panel: $8$ sites with $n=0.25$ $a_{0}^{-1}$ and $8$ wells.}\label{fig:Scaledcontactgraphs}
\end{figure}

 We find that with $\tilde{U}_{w}/\left(C_{U}t_{w}\right)=11500$ the entanglement for the QW system is almost identical to the results from the Hubbard model for all the systems considered (Fig.~\ref{fig:Scaledcontactgraphs}). 

This suggests that although the calculated $\tilde{U}_{w}/t_{w}$ gives a good estimate for  the parameter $U$ to be used in the Hubbard model, we may improve the entanglement accuracy by fitting $\tilde{U}_{w}/(C_{U}t_{w})$ and thereby compensate for some of the differences between the models.  The extent of the scaling confirms that---at least for  parameters for which there is no significant electron density in the barrier regions---the use of a single square well wave-function is a good approximation in the calculation of $\tilde{U}_{w}/t_{w}$ as the result is very close to the scaled value.

\section{Upper bound for the entanglement and strong Coulomb interaction limit}
\label{sec:chap6maxS}

Let us consider the case of zero magnetization, i.e. $P(\uparrow)=P(\downarrow)$ and $2\cal{N}$ particles where  $\cal{N}$ is an integer.  With $M$ wells, we then have a constraint from conserving the particle number 
\begin{equation}
\phi=\sum_{i}^{M} \left( P_{i}(\uparrow\downarrow)+ P_{i}(\uparrow) \right)-{\cal{N}} = 0,
\label{eq:constraint1}
\end{equation}
and constraints from the requirement that occupation probabilities for any well/site must sum to one
\begin{equation}
\psi_{i}= P_{i}(\uparrow\downarrow)+ 2P_{i}(\uparrow)+P_{i}(0) -1=0.
\label{eq:constraint2}
\end{equation}

We use Lagrange multipliers to maximize $S$ subject to these constraints, i.e.
\begin{equation}
\frac{\partial}{\partial P_{j}(\gamma)}\left (  S-\lambda \phi - \sum_{i}^{M}\mu_{i} \psi_{i} \right )=0
\end{equation}
with $\gamma=\uparrow,\uparrow\downarrow,0$.  Eliminating $\lambda$ and $\mu_{i}$ from the resulting equations give
\begin{equation}
P_{i}(\uparrow\downarrow)P_{i}(0)= (P_{i}(\uparrow))^{2}. 
\label{eq:lagrangemultiresult}
\end{equation}
Eq.~(\ref{eq:lagrangemultiresult}) relates occupation probabilities within each site, so we can find a local maximum of the entanglement where all the wells/sites are equivalent, i.e. $P_{i}(\gamma)=P(\gamma)$.  Imposing this condition on  Eqs.~(\ref{eq:constraint1}), (\ref{eq:constraint2})  and (\ref{eq:lagrangemultiresult}) gives ${P(\uparrow)={\cal{N}}/M-\left({\cal{N}}/M\right)^{2}}$ and ${P(\uparrow\downarrow)=\left({\cal{N}}/M\right)^{2}}$. We note that it is only the ratio ${\cal{N}}/M$ that matters for the probabilities and hence the entanglement.  For 2 particles,
the largest entanglement occurs for two well/sites (Fig.~\ref{fig:ContactGraphsApproxt}); this suggests that ${\cal{N}}/M=1/2$ (half-filling) may be the condition to obtain the largest maximum for the entanglement.

 We now continue with ${\cal{N}}=1$ and note that ${P(\uparrow)=1/M-1/M^{2}}$ and ${P(\uparrow\downarrow)=1/M^{2}}$ correspond to the condition of no preferred well and two uncorrelated particles of opposite spin.  

Reduced density matrices with eigenstates equal to $1/g$, $g$ the number of degrees of freedom, would correspond to maximal entanglement. However, under the stipulation of preserving the particle number together with the request that $\sum_\gamma P_i(\gamma)=1$, this state cannot be achieved except for $M=2$. We could think of moving closer to this by attempting to achieve reduced density matrices with more homogeneity within the eigenvalues, i.e., achieving $P_{i}(\gamma)\approx P_{i}(\gamma')$, at least within certain wells. We implement this by relaxing the condition that the wells are equivalent and moving part of the particle density from one site to another. We reduce $P_i(\uparrow)$ and $P_i(\downarrow)$ by $q$ while increasing  $P_j(\uparrow\downarrow)$ at site $j\ne i$ by $q$, with the empty occupation probabilities adjusted accordingly.  Setting $dS/dq=0$ gives $q=0$ suggesting that the maximum entanglement occurs when all wells are equivalent.

Under this condition, the average single-site entanglement is given by Eq.~(\ref{eq:singlesite}), and simplifies to 
\begin{equation}
S_{\text{max}}^{\text{th}}(M)=2\log_{2}(M)+2\left(\frac{1}{M}-1 \right)\log_{2}\left(M-1 \right).
\label{eq:chap6maxS}
\end{equation}
This maximum average single-site entanglement decreases as the number of wells increase  (see Table \ref{tbl:maxsinglesite}) and, for two particles, $S_{\text{max}}^{\text{th}}\stackrel{M\to\infty}{\to}0$.  In the Hubbard model picture, this would correspond to the limit of the number of sites going to infinity and  the average particle density going to zero. This limit would in fact be expected to have no entanglement as it is essentially a product state of empty occupations.

For $d=2~a_0$, $S_{\text{max}}^{\text{th}}$ is reached for $U=0$ and two wells, similarly to the Hubbard model; however, for $M>2$, some interaction is required  to balance the propensity of the non-interacting wave-function to favor inner wells.  Turning on the repulsion between electrons will tend to reduce the discrepancies between the electron density peaks in different wells (see e.g. the upper panel of Fig.~\ref{fig:Deltawelldensityd=2}); however this will also tend to decrease the double occupation probability and in particular the already too low value of $P_{1(M)}(\uparrow\downarrow)$ at the outer wells.  Therefore, due to the open boundary conditions here considered, it may not be possible for the system to reach the theoretical maximum for the entanglement by simply varying $U$, as a perfect balance between occupation probabilities in different wells may not be achieved without, for example, a spatial modulation of the particle-particle interaction.

\begin{table}
\centering
\begin{tabular}{ || c | c | c || }
\hline
  $M$ & $S_{\text{max}}^{\text{th}}$ & $S_{\text{max}}$   \\ \hline
$2$ & $2$ & $2$  \\ \hline
$4$ & $1.623$ & $1.550$ \\ \hline
$6$ & $1.300$ & $1.234$ \\ \hline
$8$ &   $1.087$ &  $1.033$     \\ \hline
\hline
\end{tabular}
\caption{Table showing the maximum theoretical average single-site entanglement Eq.~(\ref{eq:chap6maxS}), and the maximum entanglement as calculated for the QW electron system for $d=2~a_0$ and different numbers of wells.}\label{tbl:maxsinglesite}
\end{table}

\begin{table}[h]
\centering
\begin{tabular}{ || c | c | c | c ||}
\hline
$d~(a_0)$   &   $P_1(\uparrow)$ & $P_2(\uparrow)$ & $P^{th}_{\text{max}}(\uparrow)$ \\\hline
$0.2$      & $0.157$    & $0.277$           & $0.1875$      \\                        \hline
$2$ & $0.164$    & $0.265$           & $0.1875$     \\                      
\hline\hline
$d~(a_0)$ & $P_1(\uparrow \downarrow)$ & $P_2(\uparrow \downarrow)$ & $P^{th}_{\text{max}}(\uparrow \downarrow)$ \\\hline
$0.2$ & $5.62\times 10^{-3}$                  & $0.0595$ & $0.0625$\\ \hline
$2$  &   $8.58\times 10^{-3}$                 &$0.0626$ &    $0.0625$\\
\hline\hline
$d~(a_0)$ & $P_1(0)$ & $P_2(0)$ & $P^{th}_{\text{max}}(0)$ \\ \hline
$0.2$  & $0.681$ & $0.386$ &  $0.5625$   \\\hline
$2$  & $0.663$ & $0.408$  & $0.5625$    \\ 
\hline\hline
\end{tabular}
\caption{Occupation probabilities for $M=4$, interwell distances $d=0.2~a_0$ and $d=2~a_0$, and interaction value $U_{S_{max}}$ corresponding to the entanglement maximum. Theoretical values as used in Eq.~(\ref{eq:chap6maxS})}\label{tbl:d=2vsd=02}
\end{table}

In Table \ref{tbl:d=2vsd=02}  we compare the occupation probabilities $P_{\text{max}}^{\text{th}}(\gamma)$ corresponding to the maximum theoretical entanglement $S_{\text{max}}^{\text{th}}$ to the occupation probabilities calculated for the maximum value of the entanglement for the $M=4$ system and interwell distances $d=2$ and $d=0.2$. We note that the largest discrepancy with the theoretical values is observed for the double occupation probability, with $P_1(\ua\da)\ll P_{\text{max}}^{\text{th}}(\ua\da)$. This is greatly responsible for the fact that $S_{\text{max}}<S_{\text{max}}^{\text{th}}$ for both interwell distances. The $d=0.2$ system also presents the largest discrepancies  
$|P_i(\gamma)- P_{\text{max}}^{\text{th}}(\gamma)|$ for any $\gamma$ and $i$, as well as a large inhomogeneity between well occupation probabilities $|P_1(\gamma)- P_2(\gamma)|$ for  any $\gamma$. These account for the fact that this system presents a lower maximum for the entanglement in respect to the system with $d=2$ $a_{0}$. 

We may calculate the theoretical limit for the entanglement of two electrons and $M$ wells when $U\rightarrow \infty$ and all wells are equally favorable. Using a similar procedure to section \ref{sec:chap6maxS} but with $P(\uparrow\downarrow)=0$ and $P(\uparrow)=P(\downarrow)=1/M$, we obtain 
\begin{eqnarray}
\nonumber S_{U\rightarrow \infty}^{\text{th}}=\frac{2}{M}\log_{2}(M)+
\\ \left(\frac{2 }{M}-1\right)\log_{2}\left(1-\frac{2}{M} \right).
\label{eq:chap6SinfiniteU}
\end{eqnarray}
We see in Table \ref{tbl:infiniteUsinglesite} that Eq.~(\ref{eq:chap6SinfiniteU}) describes the large $U$ limit of the QW system fairly well, with a percentage error of at most 3\%. 
For $M=6$ and $M=8$  the entanglement of the QW system saturates at $U\approx40$ and remains slightly below the theoretical limiting value as in this case the assumption of equivalent wells does not hold even for very strong interactions (Fig.~\ref{fig:Deltawelldensityd=2}, lower panel). 

\begin{table}
\centering
\begin{tabular}{ || c | c | c | c || }
\hline
  $M$ &   $S_{U\rightarrow \infty}^{\text{th}}$ & $S_{U=40}$ &  $S_{U=450}$ \\ \hline
$2$ & $1$ & $1.030$  & $1.000$ \\ \hline
$4$ & $1.5$ & $1.503$ & $1.500$\\ \hline
$6$ & $1.252$ & $1.226$ & $1.226$ \\ \hline
$8$ &   $1.061$ &  $1.032$  &  $1.032$  \\ \hline
\hline
\end{tabular}
\caption{Table showing the limiting value for the average single-site entanglement entropy Eq.~(\ref{eq:chap6SinfiniteU}), and the results from the QW electron system with $U=40$ and $U=450$ for $d=2~a_0$ and different numbers of wells.}\label{tbl:infiniteUsinglesite}
\end{table} 

\begin{table}[h]
\centering
\begin{tabular}{ || c | c | c | c || }
\hline
$d~(a_0)$   &   $P_1(\uparrow)$ & $P_2(\uparrow)$ & $P^{th}(\uparrow)$ \\ \hline
$0$      & $0.196$    & $0.284$           &    $0.25$   \\\hline
$0.2$      & $0.228$    & $0.268$           &    $0.25$   \\\hline
$2$      & $0.249$    & $0.251$           &    $0.25$    \\\hline \hline
Hubbard      & $0.249$    & $0.251$           &    $0.25$     \\\hline\hline
$d~(a_0)$& $P_1(\uparrow \downarrow)$ & $P_2(\uparrow \downarrow)$ & $P^{th}(\uparrow \downarrow)$ \\\hline
 $0$ &   $2.04\times 10^{-4}$                   & $1.81\times 10^{-2}$ &  $0$   \\\hline
 $0.2$ &   $2.12\times 10^{-4}$                   & $3.46\times 10^{-3}$ &  $0$   \\\hline
 $2$          &   $6.80\times 10^{-7}$                   & $5.87\times10^{-6}$ &  $0$   \\\hline\hline
 Hubbard          &   $4.94\times 10^{-7}$                   & $4.43\times10^{-6}$ &  $0$   \\\hline\hline
$d~(a_0)$& $P_1(0)$ & $P_2(0)$ & $P^{th}(0)$ \\ \hline
$0$  & $0.608$ & $0.414$ & $0.5$  \\ \hline
$0.2$  & $0.544$ & $0.461$ & $0.5$  \\ \hline
$2$  & $0.501$ & $0.499$ & $0.5$  \\ 
\hline\hline
 Hubbard   & $0.501$ & $0.499$ & $0.5$  \\ 
\hline\hline
\end{tabular}
\caption{Occupation probabilities for $M=4$, interwell  distances $d=0$, $d=0.2~a_0$ and $d=2~a_0$, and $U=450$. The corresponding values for the Hubbard model are reported as well.
Theoretical values as used  for Eq.~\ref{eq:chap6SinfiniteU}.}\label{tbl:largeU_d=2d=0.2}
\end{table}

In Table \ref{tbl:largeU_d=2d=0.2} we explore  the differences between the theoretical limiting results and the $M=4$ system. We consider  $d=0$, $d=0.2~a_0$, $d=2~a_0$, and the results from the Hubbard model. Table \ref{tbl:largeU_d=2d=0.2}  shows that the occupation probabilities for $d=2~a_0$ are almost identical to the Hubbard model and extremely close to the theoretical limiting values. For $d=0.2$ instead, no matter how strong the Coulomb repulsion between particles is made ($U=450$ in the table),  the very narrow interwell barriers fail to counteract the effect of the boundary conditions, which favor occupation in the central wells. 
In general the inhomogeneity between well occupation probabilities $|P_1(\gamma)- P_2(\gamma)|$ increases for decreasing $d$, underlining the fact that the definition of 'sites' become more arbitrary.
However the substantially larger double occupancy probability encountered for $d<2$ increases the available degrees of freedom and hence the entanglement. This confirms the entanglement trend observed in Fig.~\ref{fig:ContactGraphsApproxt}, center panel.

\section{Attractive versus repulsive particle-particle interaction\label{attr-vs-rep}}
\begin{figure}[ht]\centering
  \includegraphics[width=.4\textwidth]{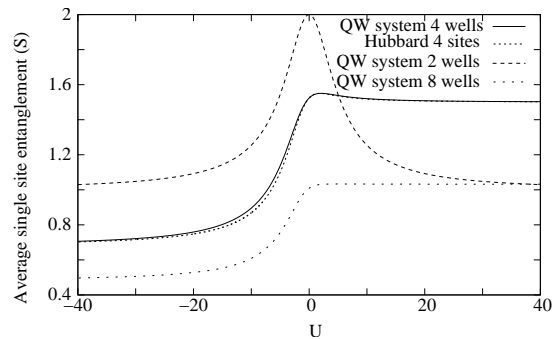}
  \caption{ Average single-site entanglement of the 4-sites Hubbard model and of the QW system with $2$, $4$, and $8$ wells vs $U$,  $U=\tilde{U}_{w}/t_{w}$, $d=2$ $a_{0}$, $w=2$ $a_{0}$, and $v_{0}=10$ Hartree.}\label{fig:varynumberofwellsScaledS}
\end{figure}

We wish to discuss how the entanglement pattern is modified when we compare attractive ($C_{U},U<0$) with  repulsive particle-particle interaction.  
In the following we will consider the QW system with $d=2~a_0$ and the Hubbard model.
In Fig.~\ref{fig:varynumberofwellsScaledS} we show the change in the average single-site entanglement with $U$ for different numbers of wells.     From our calculations $S_{\text{max}}$ always occurs for $U\ge 0$ and corresponds to $U= 0$ for two, $U= 2.1$ for four,  while $U= 4.8$ for eight wells.
As expected from Eq.~(\ref{eq:chap6maxS}), the maximum average single-site entanglement $S_{\text{max}}$  decreases with increasing number of wells and our conjectured theoretical maximum entanglement $S_{\text{max}}^{\text{th}}$ is indeed an upper bound, to which the actual system comes reasonably close (Table \ref{tbl:maxsinglesite}). For $M>2$, due to the non-periodic nature of the system, an unmodulated interaction strength drives the system towards having equivalent wells only in the very large $|U|$ limit.  However, as particle-particle interaction would naturally introduce correlations, any spatial modulation of $U$ should probably be non trivial in order to mimic the uncorrelated electrons' occupation probabilities corresponding to the maximum theoretical entanglement
Eq.~(\ref{eq:chap6maxS}).

For $U<0$ the entanglement decreases monotonically for increasing $|U|$. This is due to a disproportionate increase of the double occupation probabilities, which are favored by the attractive interparticle interaction. This limits the access to other degrees of freedom which might contribute to the entanglement, and consequently the entanglement is reduced.

Our calculations show that the Hubbard model reproduces well the average single-site entanglement of a QW system with relatively wide interwell barriers.  The comparison for $M=4$ is shown in Fig.~\ref{fig:varynumberofwellsScaledS}.

\subsection{Large inter-particle attraction limit}
For $U<<0$ the two center wells could have equal probabilities of double occupation and emptiness whilst all the other wells would be empty.  This would lead to an average single-site 
entanglement of 
\begin{equation}
S^{\text{th},1}_{U<<0}=2/M.
\label{eq:LowerBoundUNeg}
\end{equation}
We see in Table \ref{tbl:neginfiniteUsinglesite} that the QW system does not get very close to this limit except for $M=2$.  The form of the confining potential  is such that all the wells will always contain some density for the finite interaction strengths considered ($U\ge-40$).  At these interaction strengths  the system is better described by assuming that all wells are equivalent but that there is no single occupation.  This gives 

\begin{equation}
S^{\text{th},2}_{U<<0}=\frac{1}{M}\log_{2}\left (M\right)-\left(1-\frac{1}{M} \right)\log_{2}\left (1-\frac{1}{M}\right).
\end{equation}  
Fig.~\ref{fig:varynumberofwellsScaledS} shows that the entanglement remains intermediate between $S^{\text{th},1}_{U<<0}$ and $S^{\text{th},2}_{U<<0}$, due to the relatively limited effect  of the short-range interaction considered.
\begin{table}
\centering
\begin{tabular}{ || c | c | c | c || }
\hline
  $M$ & $S^{\text{th},1}_{U<<0}$ & $S^{\text{th},2}_{U<<0}$ & $S_{U=-40}$   \\ \hline
$2$ & $1$      &  $1$    & $1.030$ \\ \hline
$4$ & $0.5$    &  $0.811$    & $0.706$ \\ \hline
$6$ & $0.333$  &   $0.65$   & $0.585$  \\ \hline
$8$ &   $0.25$ &    $0.544$  & $0.497$     \\ \hline
\hline
\end{tabular}
\caption{Table showing the theoretical limits for the average single-site entanglement and $U\ll0$ for different numbers of wells. Results from the QW system with $d=2~a_0$ and $U=-40$ are presented as well.}\label{tbl:neginfiniteUsinglesite}
\end{table} 

\section{Spatial versus average single-site entanglement}
\label{sec:spatial}
In this section we consider a different type of entanglement contained within the QW system, the spatial entanglement between the two trapped particles. This represents the particle-particle entanglement spanning from  the many-body wave-function spatial degrees of freedom. Once more we calculate  the entanglement using the von Neumann entropy   of the reduced density matrix, $S_{sp}=-Tr \rho_{\text{red,sp}} \log_{2} \rho_{\text{red,sp}}$, with  $\rho_{\text{red,sp}}$ calculated from the spatial degrees of freedom as \cite{Coe}
\begin{equation}
\rho_{\text{red,sp}}({x_{1}},{x_{2}})=\int \Psi^{*}({x_{1}},{x_{3}})\Psi({x_{2}},{x_{3}}){dx_{3}}.
\end{equation}
This expression is diagonalized with respect to the basis set employed. Also in this case we allow for attractive as well as repulsive interaction between the particles.  

Notice that in the present case the spatial entanglement is zero when there is no interaction as  the wave-function factorizes into spatial and spin components and the implicit correlations arising from the Pauli exclusion principle---and the related entanglement---are accounted for within the spin degrees of freedom.

\begin{figure}[ht]\centering
  \includegraphics[width=.4\textwidth]{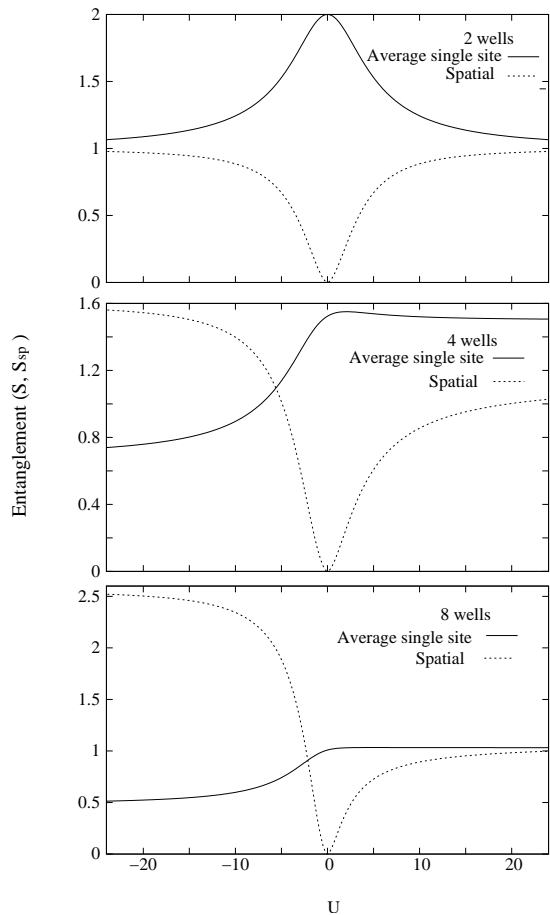}
  \caption{ Average single-site entanglement $S$ and spatial entanglement $S_{sp}$ for the QW electron system with $U=\tilde{U}_{w}/t_{w}$,   $d=2$ $a_{0}$, $w=2$ $a_{0}$, $v_{0}=10$ Hartree and $2$ wells (upper panel), $4$ wells (center panel), and $8$ wells (lower panel)}\label{fig:ScaledSpatialgraphs}
\end{figure}

For two wells (Fig.~\ref{fig:ScaledSpatialgraphs}, upper panel) we see that the spatial entanglement is a mirror image of the site entanglement when reflected along the line $S,~S_{sp}=1$. For larger numbers of wells the relationship is more complicated.  In most regions when the spatial entanglement increases the site entanglement decreases and vice versa; however the spatial entanglement does not have a minimum exactly where the average single-site entanglement has a maximum.  This is because the spatial entanglement's minimum always occurs at $U=0$ when there is no correlation between the particles' positions.  

An intuitive explanation for the almost opposite behavior of these two types of entanglement is that for $U>0$ increasing $U$ increases the repulsion and the correlation between particles.  Hence one electron's position reveals more about the other electron while the number of spatial degrees of freedom is not dramatically limited, so the spatial entanglement increases.  However the probability of double occupation is reduced by a large positive interaction so less is learned by measurement with respect to wells/sites even though the electron affects the other's position more. Therefore the site entanglement decreases once it has reached its maximum but, for $M>2$, much less strongly than the spatial entanglement's increase. 

 For $U<0$  the reduction in the probability of single occupation causes the average single-site entanglement to decrease markedly when $|U|$ increases.  The increase in spatial entanglement with increasing $|U|$ here comes from the system approaching the situation where measurement of one electron's position reveals the other electron to be in the same region.   This results in large entanglement and we find that the spatial entanglement for $U<0$ increases as the number of wells where the electron can be found increases.

\section{Conclusion}
\label{sec:Conclusions}
In this paper we examined the average single-site and spatial entanglement of two particles confined in a string of quantum wells and interacting via a contact interaction.  The results for average single-site entanglement were compared to those of the one-dimensional Hubbard model with on-site interaction, to investigate when this model is a good approximation to the two-particle system.  For repulsive (Coulomb) interaction, we found that the trend of the entanglement was reproduced, with a generally good quantitative agreement, when comparing with a Hubbard model characterized by $U=\tilde{U}_{w}/t_{w}$ where $\tilde{U}_{w}$ and $t_{w}$ were calculated from the quantum well system.  This was not entirely expected as even a contact interaction still has contributions beyond on-site interaction especially for relatively small but finite barrier widths. In the latter case the Hubbard model reproduces at least the qualitative trend, with a maximum discrepancy of $\sim 20\%$ for the parameters considered. 
We have compared the results from the  Hubbard model also with the limiting case when no barrier exists between sites and a single QW is arbitrarily divided into $M$ equal sectors, each sector corresponding to one `site'.  Surprisingly, when enough `wells' are considered, the Hubbard model reproduces the entanglement within few percent. We interpreted this as the Hubbard model sites being a fine enough mesh discretization of the continuous spatial variable.

  We conjectured a theoretical maximum value for the average single-site entanglement of two-particle trapped within $M$ wells.  We saw that the maximum value was not reached except in the case of $2$ wells.  We argued that for $M>2$ some spatially modulated particle-particle interaction is needed to reach the maximum average single-site entanglement as to counteract the propensity of the particles to occupy the inner wells.

 Despite the calculated values of $\tilde{U}_{w}$ and $t_{w}$ appearing to give very good results for relatively wide interwell barriers,  we found that an even better match between the Hubbard model and the electron system could be achieved by rescaling the value of $\tilde{U}_{w}/(C_{U}t_{w})$. This suggests that there were some small contributions to the interaction beyond the on-site repulsion for the chosen well parameters, but that the main approximation used---hopping parameters  and interaction strength independent of the site and estimated from the ground state of a single finite quantum well---remains valid. However, as the interwell barrier width decreases, these approximations fails and no rescaling of $\tilde{U}_{w}/(C_{U}t_{w})$ could improve the match between the quantum well system and the Hubbard model results. 

We also considered an attractive interaction $U<0$ and relatively wide interwell barriers. In this case  the average single-site entanglement of the quantum well system was well approximated by the Hubbard model.

Finally we have considered a different type of entanglement---the spatial entanglement---for the quantum well system. Our results showed that the spatial entanglement tends to display in most parameter regions an opposite trend in respect to the average single-site entanglement. 

Future work includes considering long range Coulomb interactions, and how this affects comparison with the results from the Hubbard model.

\end{document}